\documentclass[draft,grl]{agu2001}
\usepackage{graphicx}
\authorrunninghead{YERMOLAEV AND YERMOLAEV}

\titlerunninghead{COMMENT}

\authoraddr{Yu. I. Yermolaev
Space Plasma Physics Department, Space Research Institute, 
Russian Academy of Sciences, Profsoyuznaya 84/32, Moscow 117997, Russia. 
(yermol@iki.rssi.ru)} 
\authoraddr{M.Yu. Yermolaev
Space Plasma Physics Department, Space Research Institute, 
Russian Academy of Sciences, Profsoyuznaya 84/32, Moscow 117997, Russia.}

\begin{document}

%
%
%
%
%

%
%

\title{Comment on "Sizes and relative geoeffectiveness of interplanetary 
       coronal mass ejections and the preceding shock sheaths during intense 
       storms in 1996-2005" by J. Zhang et al.  
}

%
%



\author{Yu. I. Yermolaev and M. Yu. Yermolaev} 
\affil{Space Plasma Physics Department, Space Research Institute, 
Russian Academy of Sciences, Profsoyuznaya 84/32, Moscow 117997, Russia}

\begin{abstract}
(No abstract for comment)
\end{abstract}

%
%

%

\begin{article}

%
%

Recently 
\citet{Zhang2008} presented a statistical study of sizes and relative 
geoeffectiveness of ICMEs (bodies of magnetic clouds) and preceding 
sheaths for 46 events responsible for intense (Dst $<$ -100 nT) geomagnetic 
storms in 1996-2005 in which only a single ICME was responsible for 
generating the storm. Here we would like to comment several results 
and conclusions of this paper. 

\section*{Durations and sizes of sheath and ICME} 

\citet{Zhang2008}
found that average durations and sizes are 10.6 hr and 0.13 AU for 
sheaths and 30.6 hr and 0.37 AU for ICMEs. 

First of all it is necessary to note that the average duration of 
geoeffective sheaths is not new result. For instance, 
\citet{Huttunen2006} found that at 1 AU geoeffective sheaths has average 
duration 11 hours. Recently, 
\citet{Yermolaev2007a} have published that durations of sheaths, magnetic 
clouds (MCs) and corotating interacting regions (CIRs) generating magnetic 
storms with Dst $<$ -60 nT during 1976-2000 were 9 $\pm$ 4 (for 22 events), 
28 $\pm$ 12 (113) and 20 $\pm$ 8 (121) hours, respectively. 
We believe that the authors of paper by 
\citet{Zhang2008} should discuss the earlier published results. 

There is difference between durations of ICMEs of 0.37 AU and 0.25 AU 
obtained by 
\citet{Forsyth2006} and 
\citet{Lepping2006} and 
\citet{Zhang2008} 
discuss the possible reason of difference of their results from earlier 
published ones: 
{\it "The difference may be a selection effect due to the fact that all the 
46 events used in this study produced major geomagnetic storms, and thus 
may possess different properties, including perhaps a larger physical size 
that may help to sustain geoeffective solar wind conditions, than the general 
population of ICMEs".} 
In our paper 
\citep{Yermolaev2008} 
we estimated durations for all (independent on generating storm) events 
and they are 15.7 $\pm$ 10.1 (642 events) for sheath, 29.8 $\pm$ 20.5 (1127) 
for ejecta, 28.2 $\pm$ 13.4 (101) for MC and 20.6 $\pm$ 12.2 (718) for CIR. 
Therefore, our results show that average duration of geoeffective sheaths 
is less than one for all sheaths while durations for geoeffective MC 
(and ejecta) and all MC (and ejecta)  are the same. So, difference in ICMEs 
durations cannot be explained by study of stronger (with lower Dst index) 
storms in paper by 
\citet{Zhang2008}. 
Thus, average duration and size of ICMEs obtained by 
\citet{Zhang2008} 
are not in agreement with earlier published results and require further 
investigations and explanations. 

\section*{Relative Geoeffectiveness of Sheaths and ICMEs}

Effectiveness of any process (also as well as geoeffectiveness) is defined 
as the ratio of "output" to "input". The authors write: 
{\it "To estimate the geoeffectiveness of the sheaths and ICMEs, we use the 
well-known epsilon-parameter, which is a good proxy of the rate of energy 
input to the magnetosphere 
\citep{Akasofu1981}"}. 
They calculated epsilon-parameter and compare "total energy input": 
{\it "the total energy input during a certain period is obtained by 
integrating epsilon during the period of interest"} and finally conclude that 
{\it "The percentage due to ICMEs ranges from 2\% to 99\% with an average 
(median) value of 71\% (80\%). While ICMEs are the dominant transient 
features producing major geomagnetic storm, sheaths are also important, 
contributing about 29\% of the total energy input into the magnetosphere 
during these storms"}.

1. 
\citet{Zhang2008} 
calculated energy input 
{\it "by integrating epsilon during the period of interest"}. 
On the one hand, the derivative of Dst-index is proportional to rate of 
energy injection into the ring current 
\citep{Burton1975,OBrien2000}. 
Therefore authors should explain what physical reason requires integrating 
epsilon. On the other hand, 
it is necessary to use real duration of event for integrating epsilon.  
As we already have shown above the durations of events may be defined 
incorrectly, therefore there are serious doubts that the integrated values 
of epsilon-parameter have been counted correctly. 

2. 
\citet{Zhang2008}  
have compared only "input", but have not compared "output". This approach 
does not allow one to compare geoeffectiveness. At approximately equal 
"outputs" the obtained data can testify in favour of a hypothesis that 
sheaths are more geoeffective than ICMEs. 

3. 
The mentioned above hypothesis that sheaths are more geoeffective than 
ICMEs has been confirmed by several publications briefly discussed below. 

\citet{HuttunenKoskinen2004} 
studied 53 intense (Dst $<$ -100 nT) magnetic storms for which they could 
identify the solar wind cause. They 
{\it "classified the drivers of storms in the following categories: 
post-shock stream or sheath region (shock/sheath), magnetic cloud (MC), 
CME ejecta without the magnetic cloud structure (ejecta) and solar wind 
causes not associated with shocks or CME ejecta (other)"} 
and 
{\it "found that postshock streams and sheath regions were the most 
important storm drivers, causing nearly half (45\%) of the storms"}. 
This question also is in detail discussed in the review by 
\citet{KoskinenHuttunen2006}.

\citet{Vieira2004} 
studied the storm-time ring current evolution of 20 intense (Dst/SYM $<$ 
-100 nT) magnetic storms driven by different interplanetary structures 
from 1998 to 2001 (they classified the magnetic storm events according 
to the region in which was observed the Bz driving structure as: 
(1) sheath events; (2) magnetic cloud events; (3) corotating interaction 
regions; and (4) complex events and sub-divided the clouds into groups 
with magnetic field rotation from southward to northward (SN) or vice 
versa (NS))  and 
{\it "found that the energy injection rate is different for different 
interplanetary structures. The energy injection rate is higher for sheath 
events and lower for NS magnetic clouds. The main consequence is that 
the magnetosphere assumes different configurations depending on the 
energy injection rate, leading to a different evolution and decay of 
the symmetric and the partial ring current".}

In our papers 
\citep{Yermolaev2007a,Yermolaev2007b} 
we studied 22 sheaths and 113  magnetic clouds during 1976-2000 resulting 
in magnetic storms with Dst $<$ -60 nT and found that 
{\it "Though the lowest values of the Bz-component of the IMF are observed 
in the MC, the lowest values of the Dst-index are achieved in the Sheath. 
Thus, the strongest magnetic storms are induced during the Sheath rather 
than during the MC body passage, probably, owing to higher pressure in 
the Sheath."} 
We have also indicated that higher geoeffectiveness of sheath may be 
connected with higher level of variations of density and magnetic field 
in the sheaths.

Thus, without discussion earlier obtained results the paper by 
\citet{Zhang2008} 
presents only limited interest as their results do not allow one 
to see a full physical picture in the discussed area of the  modern science.

\begin{acknowledgements} 
       Work was in part supported by RFBR, grant 07-02-00042.  
\end{acknowledgements}

\end{article}


\begin{thebibliography}{}

   \bibitem[{\it Akasofu}(1981)]{Akasofu1981}
\reference
Akasofu, S.-I. (1981), Energy coupling between the solar wind and the 
magnetosphere, Space Sci. Rev., 28, 121.

   \bibitem[{\it Burton et al.}(1975)]{Burton1975}
\reference
Burton, R.K., McPherron, R.L., Russell, C.T., (1975) An empirical 
relationship between interplanetary conditions and Dst. 
J. Geophys. Res, 80, 4204-4214.

   \bibitem[{\it Forsyth et. al.}(2006)]{Forsyth2006}
\reference
Forsyth, R. J., et al. (2006), ICMEs in the inner heliosphere: Origin, 
Evolution and propagation effects, Space Sci. Rev., 123, 383, 
doi:10.1007/s11214-006-9022-0.

   \bibitem[{\it Huttunen and Koskinen}(2004)]{HuttunenKoskinen2004}
\reference
Huttunen K. E. J. and Koskinen H. E. J. (2004), Importance of post-shock 
streams and sheath region as drivers of intense magnetospheric storms 
and high-latitude activity, Annales Geophysicae 22: 1729-1738

   \bibitem[{\it Huttunen et. al.,}(2006)]{Huttunen2006}
\reference
Huttunen, K. E. J., H. E. J. Koskinen, A. Karinen, and K. Mursula (2006), 
Asymmetric development of magnetospheric storms during magnetic clouds 
and sheath regions, Geophys. Res. Lett., 33, L06107, doi:10.1029/2005GL024894.

   \bibitem[{\it Koskinen and Huttunen}(2006)]{KoskinenHuttunen2006}
\reference
Koskinen, H. E. J.; Huttunen, K. E. J. (2006) Geoeffectivity of Coronal 
Mass Ejections, Space Science Reviews, Volume 124, Issue 1-4, pp. 169-181

   \bibitem[{\it Lepping et. al.,}(2006)]{Lepping2006}
\reference
Lepping, R. P., D. B. Berdichevsky, C.-C. Wu, A. Szabo, T. Narock, 
F. Mariani, A. J. Lazarus, and A. J. Quivers (2006), A summary of Wind 
magnetic clouds for years 1995 - 2003: Model-fitted parameters, associated 
errors and classifications, Ann. Geophys., 24, 215.

   \bibitem[{\it O'Brien and McPherron}(2000)]{OBrien2000}
\reference
O'Brien, T.P., R.L. McPherron (2000)  Forecasting the ring current 
index Dst in real time,  Journal of Atmospheric and Solar-Terrestrial 
Physics 62 1295-1299

   \bibitem[{\it Vieira et. al.,}(2004)]{Vieira2004}
\reference
Vieira L. E. A., Gonzalez W.D., Echer E. and Tsurutani B. T.(2004), 
Storm-intensity criteria for several classes of the driving interplanetary 
structures, , Solar Physics 223: 245-258

   \bibitem[{\it Yermolaev et. al.,}(2007a)]{Yermolaev2007a}
\reference
Yermolaev, Yu.I., Yermolaev, M.Yu., Nikolaeva, N.S., Lodkina, I.G. (2007a) 
Interplanetary conditions for CIR-induced and MC-induced geomagnetic storms, 
Bulgarian J.Phys., v.34, N 2, 2007, 128-135 

   \bibitem[{\it Yermolaev et. al.,}(2007b)]{Yermolaev2007b}
\reference
Yermolaev, Yu. I., M. Yu. Yermolaev, I. G. Lodkina, and N. S. Nikolaeva 
(2007b), Statistical Investigation of Heliospheric Conditions Resulting 
in Magnetic Storms, ISSN 0010-9525, Cosmic Research, 2007, Vol. 45, No. 1, 
pp. 1-8., Pleiades Publishing, Ltd., 2007. Original Russian Text, 
Yu.I. Yermolaev, M.Yu. Yermolaev, I.G. Lodkina, N.S. Nikolaeva, 2007, 
published in Kosmicheskie Issledovaniya, 2007, Vol. 45, No. 1, pp. 3-11.
 
   \bibitem[{\it Yermolaev et. al.,}(2008)]{Yermolaev2008}
\reference
Yermolaev, Yu. I., M. Yu. Yermolaev, I. G. Lodkina, and N. S. Nikolaeva 
(2008), Catalog of large-scale phenomena of solar wind during 1976-2000, 
Cosmic Research, Vol. 46, No. 5, (in press)

\bibitem[{\it Zhang et al.}(2008)]{Zhang2008}
\reference
Zhang, J., W. Poomvises, and I. G. Richardson (2008), Sizes and relative 
geoeffectiveness of interplanetary coronal mass ejections and the preceding 
shock sheaths during intense storms in 1996-2005, Geophys. Res. Lett., 35, 
L02109, doi:10.1029/2007GL032045.



\end{thebibliography}
\end{document}